# Silicon MOS Pixel Based on the Deep Trapping Gate Principle: Prospects and Challenges

Nicolas T. Fourches[a], Gabrielle Regula[b], Wilfried Vervisch[b]

## I. Introduction, Principles and Goals

High point to point resolution inner vertex detectors with improved radiation hardness will be needed in high energy experiments. Pixels detectors technologies include the CMOS[1] sensors [1], the DEPFET[2] [2], HPD[3]. Despite their very high spatial resolution, present day CMOS sensors are sensitive to bulk radiation damage (e.g. neutrons). LHC[4] upgrades or Super-LHC will impose them increased hardness if there are taken as a solution for vertexing. A new device was recently proposed [3-4] with the goal of decreasing pixel size and improved radiation tolerance. This pixel is based on the Deep Trapping Gate (DTG) principle. With this operation mode the 3T (transistor) CMOS pixel can be reduced to 1T. The charge harvesting diode of the standard CMOS pixel is replaced by a new kind of buried gate (the Deep Trapping Gate) which selectively traps charge carriers (holes in this case). Its mode of operation involves the presence electron or hole- deep-levels in the bandgap of semiconductor or the presence of a quantum well, which retains the carriers for a duration matching the readout time scale. A simple analysis shows that deep level density should be larger than $10^{18}$ cm$^{-3}$. This feature may prevent any major operating degradation after high neutron fluences ($10^{16}$ cm$^{-2}$) irradiation as device simulations (TCAD) show. The DTG selectively confines holes or electrons into an internal layer located closer than one micron from the channel. The implantation of neutral or active impurities providing trapping centers at defect or impurity sites is the first obvious way through. A Quantum Dot Trapping Gate with a germanium-silicon zone obtained by epitaxial growth or ion implantation followed by annealing provides an alternative solution. In the same way as the external gate, the field effect due to the DTG modulates the source to drain current, enabling ionizing particle detection through the readout of the magnitude of the source or drain current. Device shrinkage is also possible. Hole-trap introduction by deep impurity implant is the first development option, because it can be straightforwardly derived from mainstream CMOS processes. The associated band-diagram is represented in Fig. 1. In the case of n-channel device the quantum dot/box (QDTG: type II quantum dot trapping gate) [6], due to hole localization, which results in the drain-source current modulation.

[a]CEA Saclay, IRFU/SEDI, 91191, GIF/YVETTE, FRANCE
[b]Aix-Marseille Université, CNRS, IM2NP, UMR 7334, 13397, Marseille, France

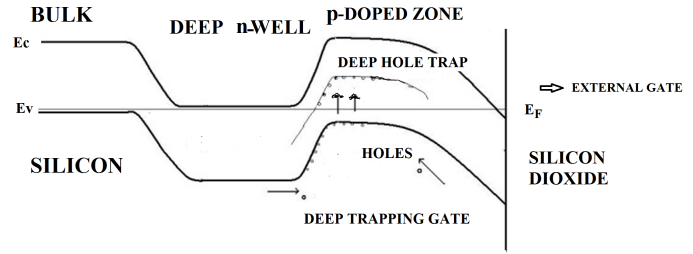

Fig. 1: Band-diagram of the trapping MOS structure showing the deep trapping gate. The cut is perpendicular to the structure (n-channel). A slight positive bias is assumed on the upper gate on the right side of the structure.

The field effect buried gate can be used for micron or sub-micron scale transistors operating at room temperature as the holes are localized almost 100 meV above the valence band edge. Then the quantum dot (or box for larger device size) provides a sufficiently high retention time for holes. Fig. 2 is the corresponding band-diagram for a QDTG with a germanium-enriched dot ($Si_{1-x}Ge_x$, $x\sim1$). This latter fabrication way is an alternative if impurity or impurity defect control shows up to be difficult with the first development option the objectives of which are the controlled introduction of deep hole traps with a concentration of electron traps as low as possible. The bottlenecks are related to material processing needed for the full manufacturing the DTG device. They both need a deep-n-well to be functional. For the two principles (deep-levels or quantum box) fabrication based on ion implantation or epitaxy can be proposed. Electrical and optical characterization of the material/device is necessary at this step together with structural material analysis.

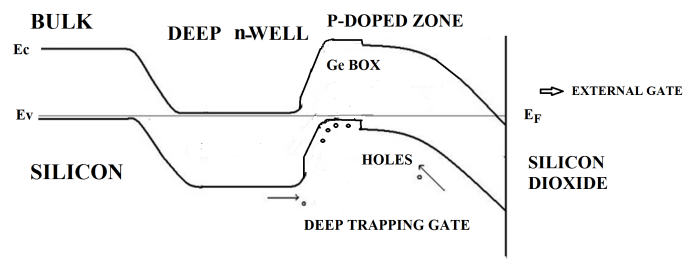

Fig. 2: Band-diagram of the trapping MOS structure with a Ge quantum box structure as a deep trapping gate, this box reduces to a dot when downsizing. The same conditions as Fig. 1 are assumed.

## II. Processing the Device: Challenges

### 1. Ion Implantation

For an n-channel transistor a buried deep n-well is necessary for the operation of a thin device (less than 2 microns). This well can be fabricated by P or As implantation at high energy (> 200 keV) into the silicon substrate followed by thermal annealing as in standard CMOS processes. The fabrication of the impurity implanted DTG is more difficult. Some deep-level studies [7] indicate that Zn induces two levels in the silicon bandgap one at $E_v + 0.60$ eV, the other at $E_v + 0.27$ eV which are hole traps. They could be used as a deep acceptor within the trapping gate. However simulations (SUPREM like TCAD tools) cannot be used to obtain the concentration of implanted electrically active deep defects, because defects reactions are not fully quantified. Zn implantation concentration should reach $10^{19}$ cm$^{-3}$ and some thermal budget should follow. SRIM simulations show that the total introduction rates are $14 \times 10^4$ cm$^{-1}$ for the Zn ions, and $4 \times 10^4$ cm$^{-1}$ for the phosphorous (P) ions (at 100 keV and 300 keV respectively, for 2500 up to more than 10000 ions events in the Si substrate, DTG depth: 0.1 μm). For a p-type background doping level of $10^{17}$cm$^{-3}$, the implantation dose needed to create an n-type well should be of the order of $5 \times 10^{12}$cm$^{-2}$. Reaching a peak of $10^{19}$ cm$^{-3}$ require a Zn implantation dose above $1 \times 10^{14}$cm$^{-2}$. This dose is lower than the critical dose for amorphization estimated with the Critical Damage Energy Density model (CDED, [8] for Ge, [9] for Si). Moreover TCAD simulations show that practically all ions should be electrically active. To be accurate amorphization requires an energy release in the silicon crystal lattice of the order of 25 eV per atom. SRIM simulations show that the loss of energy of 100 keV ions is mainly made through phonon emission and ionization. If we consider the Zn 100keV ions implanted in silicon, the figures show that implantation doses set below $5 \times 10^{13}$ cm$^{-2}$ avoid any possible amorphization. This leads to a peak concentration of Zn in the $10^{18}$-$10^{19}$cm$^{-3}$ range. These limits are still compatible with the values imposed by the operation of the DTG. The fabrication of the quantum dot/box by ion implantation should difficult as the Ge dose should be set above $3 \times 10^{17}$ cm$^{-2}$(a Ge peak concentration above $5 \times 10^{22}$ cm$^{-3}$). At these doses SRIM (ions + recoils) simulations show that peak energy of 3600 eV per silicon atom is deposited. The implantation/anneal procedure necessary to avoid amorphization constitutes then an important issue. In any case it is preferable to make the DTG in a first fabrication step if the subsequent anneals and oxidizations steps do not prevent this. The lateral distribution of Zn/Ge ions sets a side limit impacting the device's dimensions. Lateral defect control induced by the ion implantation may also be difficult to handle.

---
[1] complementary oxide semiconductor
[2] depleted p-channel field effect transistor
[3] hybrid pixel detectors
[4] large hadron collider

## 2. Epitaxial Growth

To overcome the previously described limitations, epitaxial growth widely used in microelectronic processes may be one option (Fig.3). For a zinc-doped ($10^{18}$cm$^{-3}$) DTG a homoepitaxial silicon layer is the starting point. No lattice mismatch should occur in this case. The maximum solubility of the Zn and the way to introduce Zn containing molecules could set some limits to the technique.

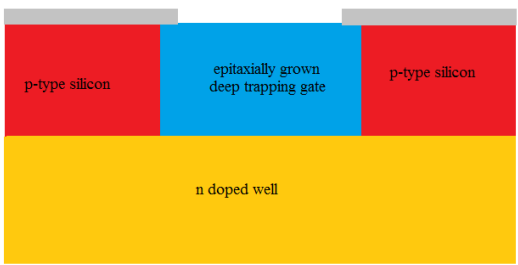

Fig. 3: Silicon substrate with an epitaxially grown doped or dot/box deep trapping gate. Etching the Si layer through a mask (grey) can be used to groove the substrate.

This technique circumvents the need to anneal the device in order to reduce defect density. The upper gate oxidation stage takes place in the last steps. For the (Si$_{1-x}$Ge$_x$, x~1) QDTG hetero-epitaxy may induce a strained and generate extended defects that could be detrimental to device operation. This problem will be addressed in the light of current fundamental knowledge. To avoid it, the thickness of the Ge layer should be reduced, with no marked effect on device operation. To limit the lateral spread of impurity and defects, selective etching could be used to reduce the lateral size of the epitaxial DTG as well as in the ion implanted one.

## 3. Readout Scheme

The device is a single pixel used in arrays. The drain is connected to a positive voltage and the bulk is grounded. All the sources are tied to a column current bias and the upper gates are connected to a row control. In detection mode, the upper gate is negatively biased so that the holes generated in the bulk by a particle migrate towards the DTG. No current flow through the device. In the read mode the upper gate is positively biased so that a current flows through the device. For readout the source voltage can be amplified and measured (Fig. 4). This makes this pixel less space consuming. A binary readout scheme has been proposed with In Pixel circuitry [10].

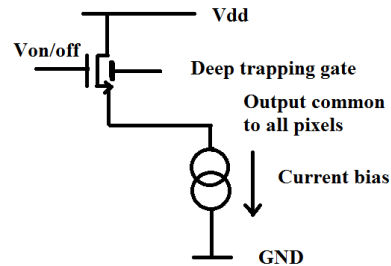

Fig. 4: Pixel schematic using the n-MOS device described above. The bulk contact is grounded and the trapping gate contact left floating. A bulk contact can be used for DTG reset.

## III. Outlook and Conclusions

We have introduced and discussed different fabrication techniques to develop a process for the DTG pixel detector. A careful preliminary analysis has help to the finding of many bottlenecks. Basic material studies should then be made. The fabrication of the DTG or (Trapping MOS) pixel can based on existing techniques in spite of the issues of downscaling to submicron scale. pixel is should be simpler to operate than former 3T CMOS sensors. It is compatible with a partially depleted operation mode.